\documentclass[prc,twocolumn,superscriptaddress,nofootinbib,floatfix]{revtex4-1}
\usepackage{epsfig}
\usepackage{graphicx}
\usepackage{palatino}
\usepackage[english]{babel}
\usepackage{hyphenat}
\usepackage{amsmath}
\usepackage{amssymb}
\usepackage{mathtools}
\usepackage{mathrsfs}
\usepackage{slashed}
\usepackage{epstopdf}
\usepackage{xcolor}
\usepackage{braket}
\usepackage{booktabs}
\definecolor{lcolor}{rgb}{0.,0.0,0.}
\definecolor{citcolor}{rgb}{0,0.,0.5}
\usepackage[breaklinks,colorlinks,urlcolor=blue,citecolor=blue,linkcolor=blue]{hyperref}
\usepackage{multirow}
\usepackage{ltablex}
\usepackage{soul}

\newcommand{\beq}{\begin{eqnarray}}
\newcommand{\eeq}{\end{eqnarray}}

\newcommand{\bem}{\begin{multline}}
\newcommand{\eem}{\end{multline}}
\newcommand{\beg}{\begin{gather}}
\newcommand{\eeg}{\end{gather}}

\newcommand{\nn}{\nonumber\\}

\newcommand{\ben}{\begin{eqnarray*}}
\newcommand{\een}{\end{eqnarray*}}

\newcommand{\secn}[1]{Section~1}
\newcommand{\appn}[1]{Appendix~1}

\long\def\comment#1{ }

\def\and{\quad\text{and}\quad}

\def\0{{\boldsymbol 0}}

\def\k{{\boldsymbol k}}

\def\0{{\boldsymbol 0}}

\begin{document}

\title{Picturing QCD jets in anisotropic matter:\\ from jet shapes to Energy Energy Correlators}

\author{João Barata}
\email[]{jlourenco@bnl.gov}
\affiliation{Physics Department, Brookhaven National Laboratory, Upton, NY 11973, USA}
\author{José Guilherme Milhano}
\email[]{gmilhano@lip.pt}
\affiliation{LIP, Av. Prof. Gama Pinto, 2, P-1649-003 Lisboa, Portugal}
\affiliation{Departmento de F{\'{i}}sica, Instituto Superior T{\'{e}}cnico (IST), Universidade de Lisboa, Av. Rovisco
Pais 1, P-1049-001 Lisboa, Portugal}
\author{Andrey V. Sadofyev}
\email[]{andrey.sadofyev@usc.es}
\affiliation{Instituto Galego de F{\'{i}}sica de Altas Enerx{\'{i}}as,  Universidade de Santiago de Compostela,\\ Santiago de Compostela 15782,  Spain}

\begin{abstract}
Recent theoretical developments in the description of jet evolution in the quark gluon plasma have allowed to account for the effects of hydrodynamic gradients in the medium modified jet spectra. These constitute a crucial step towards using jets as tomographic probes of the nuclear matter they traverse. In this work, we complement these studies by providing leading order calculations of widely studied jet observables, taking into account matter anisotropies. We show that the energy distribution inside a jet is pushed towards the direction of the largest matter anisotropy, while the away region is depleted. As a consequence, the jet mass and girth gain a non-trivial azimuthal dependence, with the average value of the distribution increasing along the direction of largest gradients. However, we find that, for these jet shapes, matter anisotropic effects can be potentially suppressed by vacuum Sudakov factors. We argue that the recently proposed measurements of energy correlations within jets do not suffer from such effects, with the azimuthal dependence being visible in a large angular window, regardless of the shape of the distribution.
\end{abstract}

\maketitle

%%%%%%%%%%%%%%%%%%
\section{Introduction}
\label{sec:intro}
%%%%%%%%%%%%%%%%%% 
Over the last decades, jets have provided clear evidence for the production of the quark gluon plasma (QGP) in high-energy heavy-ion collisions (HICs) at RHIC and the LHC. Early experimental measurements revealed that the nuclear modification factor measured from jets was significantly suppressed for intermediate $p_t$ jets, signaling the emergence of collectivity associated to the new exotic state of matter~\cite{PHENIX:2001hpc,STAR:2002ggv,ALICE:2015mjv,CMS:2016uxf,ATLAS:2018gwx,STAR:2020xiv}. In more recent years, there has been a push towards more differential studies of jets in HICs. Particular attention has been paid to understanding the details of the angular structure and real time fragmentation of partons in the QGP, and how such modifications relate to the medium properties, see~\cite{Mehtar-Tani:2013pia, Andrews:2018jcm, Cao:2020wlm,Apolinario:2022vzg} for recent reviews and further references.

Among the several recent theory developments, a lot of focus has been put on the description of jets in non-trivial backgrounds.  Examples of such efforts include the study of the influence of the early stages of HICs and of the presence of flowing matter in the QGP phase on jet properties, which have been discussed in a series of recent works~\cite{Armesto:2004pt,Sievert:2018imd,Ipp:2020nfu,Sadofyev:2021ohn,Carrington:2021dvw,Antiporda:2021hpk,Sadofyev:2022hhw,Barata:2022wim,Andres:2022ovj,Fu:2022idl,He:2020iow,Andres:2022ndd,Hauksson:2023tze,Boguslavski:2023alu}.
In parallel to these efforts, some of us and collaborators have recently provided a broad and complete theoretical description of parton dynamics in a QGP fireball, taking into account the presence of hydrodynamic gradients in the matter~\cite{Sadofyev:2021ohn,Sadofyev:2022hhw,Barata:2022krd,Barata:2022utc,Barata:2023qds}. So far, the discussion of these new effects in the evolution of the jet's partons has been framed at the level of quantities which cannot be experimentally measured but are easier to compute on the theory side. Since the ultimate goal of the jet tomography program
is to provide a complete differential description of the QCD matter produced in HICs through jet observables, in this paper we take the first step towards bridging the gap between theory and phenomenology.

\begin{figure}[h!]
  \centering
  \includegraphics[width=.45\textwidth]{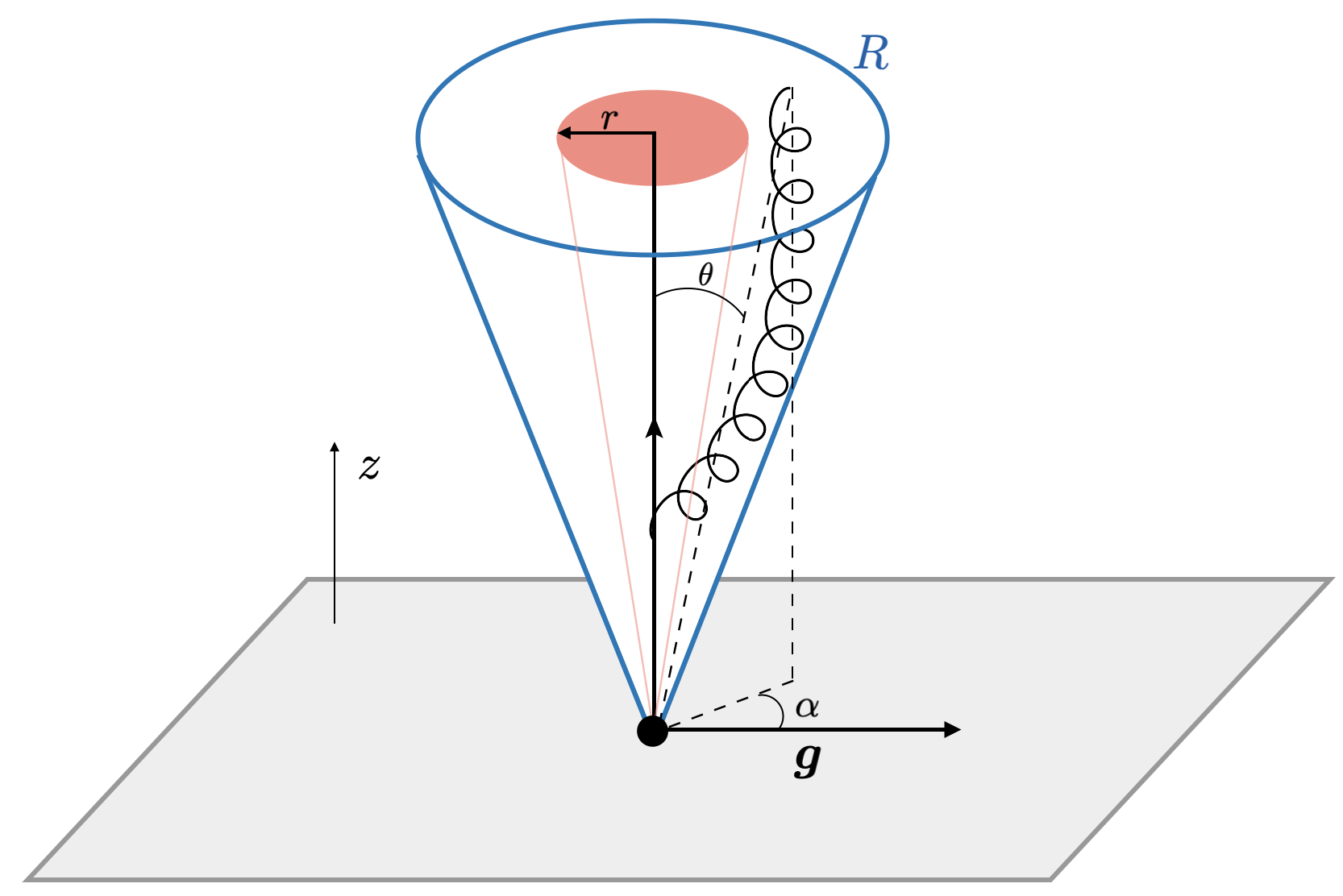}
  \caption{Diagrammatic depiction of the setup considered to study jet evolution in an anisotropic medium. The gray rectangle represents the background, where $\boldsymbol{g}$ denotes the direction along which the matter gradients are aligned. The medium is assumed to be static. The jet is composed of a quark and a single gluon. The dynamics is constrained to the plane transverse to the original momentum of the jet, aligned along $z$. The angles $\alpha$ and $\theta$ denote the angular separation between the gluon and $\boldsymbol{g}$ in the transverse plane and the gluon and the quark, respectively. The jet is assumed to be reconstructed with a jet radius $R$.}
  \label{fig:JS}
\end{figure}

To this end,  we consider the evolution of a high-energy partonic shower in a static brick of QGP matter with the initial momentum of the leading parton being aligned along $z$,  see Fig.~\ref{fig:JS}.  Matter anisotropies are introduced via the two dimensional vector\footnote{Note that the gradients are three dimensional vectors,  but only their transverse projection with respect to the jet axis is relevant \cite{Sadofyev:2021ohn}.} $\boldsymbol{g}$, along which the matter density $n$ and Debye mass $\mu$ have their largest gradients. Although this matter model is far from a realistic description of the structure of hydrodynamic gradients in a QGP fireball, it allows for a closed semi-analytical treatment, as we show below.  Going beyond this effective description in a significant way requires a more complex modeling of the matter.  One would have to build a tool including a realistic hydrodynamic profile of the QGP and a jet quenching Monte Carlo generator,  implementing the theoretical developments introduced in~\cite{Barata:2023qds}.  The numerical machinery required for this extension does not currently exist.\footnote{See~\cite{Antiporda:2021hpk} for a related study attempting to close this gap.}

Within this simple matter model, we consider the fragmentation of a hard quark in the matter at leading order (LO) in the strong coupling constant $\alpha_s$. We then compute several jet observables on the final particle distribution: the jet shape density, the lowest order jet angularities, and the double differential Energy Energy Correlator (EEC). We chose this set of observables for  several reasons. The jet shape was studied in the past in the early theoretical models including the presence of the QGP flow~\cite{Armesto:2004pt}. Jet angularities form a complete set of jet observables and their structure is well understood in vacuum QCD~\cite{Gallicchio:2012ez, Almeida:2008yp,Berger:2003iw,Reichelt:2021svh}.\footnote{For recent studies in HIC context see~\cite{Andrews:2018jcm,ALICE:2021njq,Budhraja:2023rgo}.} Also, they are among the simplest observables sensitive to jet substructure. Finally, energy flow correlations between angular separated regions in high-energy scattering events have been recently considered as a new window to study the structure of jets in vacuum~\cite{Lee:2022ige, Craft:2022kdo} and in different types of nuclear matter~\cite{Andres:2022ovj,Liu:2023aqb,Devereaux:2023vjz}, see also~\cite{Basham:1978bw,Basham:1978zq}. As we will show, they have features which might be relevant to study non-trivial modifications to the azimuthal angular structure of jets evolving in the QGP. 

Our study has several limitations which we follow to enumerate. As already mentioned, we take a simplistic model for the matter, which allows for a controlled theoretical calculation. Despite this fact, the setup allows to qualitatively assess the modifications to the jet due to the presence of matter gradients. As also mentioned previously, our calculations, for most observables, are done at LO in the strong coupling constant, i.e. we only consider a single gluon being produced from the original hard quark. Going beyond LO is, at this moment, extremely challenging from the theoretical point of view~\cite{Fickinger:2013xwa,Arnold:2023qwi}. Nonetheless, one should not expect LO (or other fixed higher order) calculations to give quantitative or qualitatively accurate descriptions of jet observables. Historically, this barrier has been surpassed by employing phenomenological inspired models~\cite{Mehtar-Tani:2021fud, Caucal:2021cfb} or by using Monte Carlo codes~\cite{Zapp:2008gi, Casalderrey-Solana:2014bpa,Caucal:2019uvr}, which can mimic some important higher order effects. In this work we do not employ such strategies for two reasons. First, as mentioned above, no Monte Carlo code is currently available that implements the theoretical effects we consider here. Second, employing phenomenological models will drive our calculation further way from the controlled theory framework being employed, blurring the line between modeling and a first-principle calculation. In effect, we attempt to give priority to the latter. Finally, we work in the limit where the gluon is always softer than the quark, which might neglect important sectors of the full phase space. We will comment further on this limitation for particular observables.

This manuscript is organized as follows: section~\ref{sec:jet shape} presents the calculation of the  jet shape density, section~\ref{sec:jet angularities} introduces the calculation of the lowest order jet angularities at leading logarithmic accuracy in the medium, and finally in section~\ref{sec:EEC} we discuss how the EECs in a jet can be used to study matter gradients. Our conclusions are detailed in section~\ref{sec:conclusion}.

%%%%%%%%%%%%%%%%%%
\section{Jet shape density}
\label{sec:jet shape}
%%%%%%%%%%%%%%%%%%

We consider first the effects of medium anisotropies on the energy distribution inside a jet.  Along with the multiplicity distributions,  see e.g. \cite{Salgado:2003rv},  this was one of the first observables to be theoretically computed for medium modified jets in models with non-trivial backgrounds~\cite{Armesto:2004pt}. To this end, we define the jet shape as
\begin{align}
 \rho(r) = \int_0^r  dr' \, \frac{p_t(r')}{p_t^{\rm jet}}\, ,
\end{align}
where $p_t^{\rm jet}$ denotes the total transverse momentum of the jet in detector coordinates, and $r =\sqrt{\Delta \phi^2 + \Delta \eta^2} $ is defined as the radial displacement with respect to the jet axis in $(\phi,\eta)$ coordinates. In this study, we always assume that partons evolve at midrapidity ($\eta\approx 0$) and consider small opening angle for the jets. Under these assumptions, one can identify $r$ with the angular distance with respect to the jet axis and $p_t^{\rm jet}$ with the total energy of the jet, both measured in the local jet frame. Therefore, the jet shape $\rho(r)$ accounts for the amount of energy contained in a cone of radius $r$ inside a larger jet with radius $R>r$, see Fig.~\ref{fig:JS}.

At LO in $\alpha_s$ and taking into account only medium-induced effects,  it is simple to show that~\cite{Armesto:2004pt}
\begin{align}\label{eq:JS_1}
\rho(r)&= 1 - \frac{1}{p_t^{\rm jet}} \int_0^{p_t^{\rm jet}} d\omega \int_{\omega r }^\omega   d^2\k \, \omega \frac{dI}{d\omega d^2\k}   \, ,
\end{align}
where the maximum jet radius is set to $R=1$,  assuming that the dynamics is dominated by the collinear modes, and the integration limits on the $d^2\k$ integral are for its radial part. The purely medium-induced radiation spectrum $dI$,  sensitive to the QGP anisotropies,  depends on the gluon energy $\omega<p_t^{\rm jet}$ and the gluon transverse momentum $\k$,  satisfying $k<p_t^{\rm jet}$.  This spectrum also depends on the Debye mass $\mu$ and density of color sources $n$. Following~\cite{Sadofyev:2021ohn,Barata:2022krd,Barata:2023qds}, we take into account the medium structure by employing the hydrodynamic gradient expansion at the level of $dI$. To leading order in gradients, one can write the spectrum as~\cite{Barata:2023qds}
\begin{align}\label{eq:spec}
  \frac{dI}{d\omega d^2\k}  = \frac{dI_0}{d\omega d^2\k} + \hat {\mathbf{g}} \cdot \k \frac{dI_1}{d\omega d^2\k} +\mathcal{O}\left(\hat { \mathbf{g}}^2\right)\, ,
\end{align} 
while accounting for all possible gluon exchanges between the medium and the jet. The leading ($I_0$) and subleading order ($I_1$) contributions to the radiation spectrum are detailed in~\cite{Barata:2023qds}. In turn, $\hat { \mathbf{g}}$ is a two-dimensional vector operator linear in the medium gradients. Below, in order to numerically compute the gluon spectrum we make use of the ubiquitous harmonic approximation for the scattering potential in the medium, under which $\hat { \mathbf{g}}$ reduces to a single gradient vector,  see Fig.~\ref{fig:JS}, and thus we drop the operator notation.  This gradient vector is defined by $ \mathbf{g}\cdot \k \approx 3 \frac{|\boldsymbol{\nabla} T|}{T} k \cos\alpha\equiv 3 \gamma_T T k \cos\alpha$, where $k\equiv |\k|$,  $\alpha$ is the angle between $\k$ and $\mathbf{g}$, and we have introduced a shorthand notation $\gamma_T$ for further convenience.  We have introduced $T$ as the medium temperature, and assumed that $n\sim T^3$ and $\mu\sim g T$ scale uniformly with $T$,  see~\cite{Barata:2023qds} for details. We denote the transverse temperature gradients by $\boldsymbol{\nabla} T$, which is a two dimensional vector in the plane transverse to the jet axis. Finally, the medium-induced term can be related to a normalized cross-section by\footnote{This can be already seen at the level of the vacuum cross-section,  {\it c.f.} Eq. (6) in~\cite{Andres:2022ovj} and Eq. (3.2) in~\cite{Barata:2021wuf}.} 
\begin{align}
	\frac{d\sigma}{\sigma dx d\theta d\alpha} = \omega \frac{dI}{d\omega d^2 \k} p_t^{\rm jet} k\, ,
\end{align}
where we have introduced the gluon energy fraction $\omega \equiv x p_t^{\rm jet}$ and the polar angle $\theta \equiv k/\omega$.

\begin{figure*}[htbp]
  \centering
  \includegraphics[width=.9\textwidth]{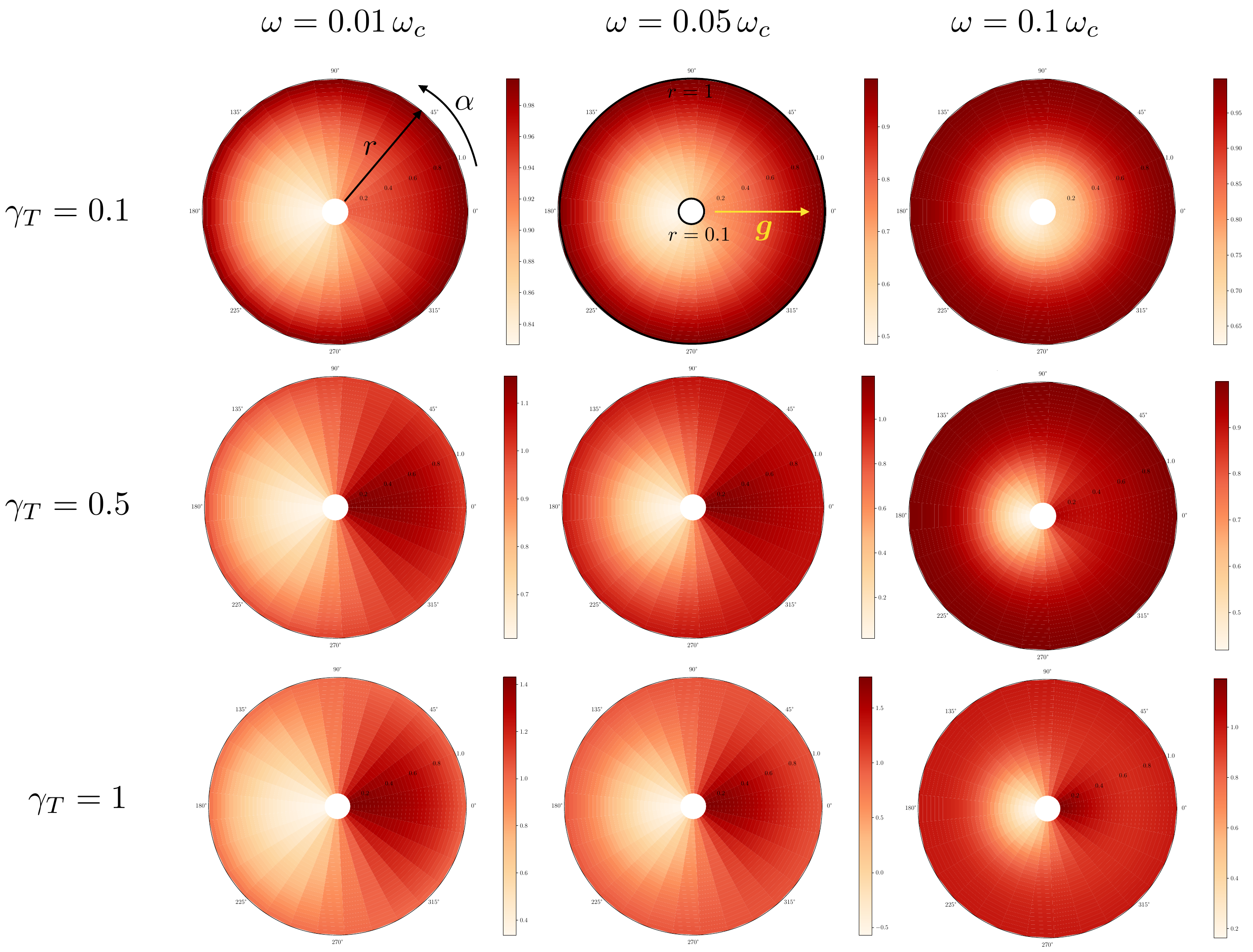}
  \caption{Jet shape density profiles for the following set of parameters: $p_t^{\rm jet}=100$ GeV, $\hat q =0.2 \, {\rm GeV}^2 {\rm fm}^{-1}$, $L=5$ fm, $T=300$ MeV,  and $\alpha_s=0.28$.  The top, middle, and bottom rows correspond to $\gamma_T=0.1$, $\gamma_T=0.5$,  and $\gamma_T=1$.  These values cover an ample and phenomenologically relevant range, see \cite{Barata:2023qds} for $\gamma_T$ estimates.  The columns, going from left to right, correspond to $\omega= 0.01\,\omega_c$, $\omega= 0.05\, \omega_c$,  and $\omega= 0.1\, \omega_c$, where we have defined the critical frequency as $\omega_c = \frac{1}{2}\hat q L^2$. }
  \label{fig:JS_1}
\end{figure*}

Evaluating Eq.~\eqref{eq:JS_1} is numerically demanding since it requires accounting for all possible gluons within a jet cone. In addition, the observable is inclusive in azimuthal angle which averages out the medium modifications we are interested in here.  For this reason,  we consider instead the jet shape density distribution
\begin{align}\label{eq:dens_JS}
  (2\pi) p_t^{\rm jet}\frac{d\rho(r)}{d\omega d\alpha}&=   1  -  2\pi \int_{\omega r}^\omega   dk k \,  \omega \frac{dI}{d\omega d^2\k}\, .
\end{align}
In this formulation, $d\rho(r)$ measures the contribution to the jet energy due to the emission of gluons of energy $\omega$ at angle $\alpha$. The numerical evaluation of Eq.~\eqref{eq:dens_JS} for several values of $\omega$ and $\gamma_T$ is provided in Fig.~\ref{fig:JS_1}. As expected, increasing the value of the temperature gradient through $\gamma_T$ (top to bottom in each column) the energy distribution inside the jet becomes increasingly more asymmetric, with most radiation aligning along the anisotropy direction ($\alpha=0$). In addition, gradient effects are more prominent for softer gluons (left most column), while for the most energetic gluons the distribution remains fairly isotropic for large enough values of $r$. Interestingly, we observe in most plots that the maximum of the density profile is not achieved at the boundary, as in the vacuum, but rather in an intermediate radial region. This peak is compensated on the away side by an energy valley. As a result, when computing a directional observable with a non-trivial transverse dependence, i.e. with a profile along the radial direction in Fig.~\ref{fig:JS_1}, the associated distribution can have a complex dependence on the gluon energy. A related observation had already been considered in~\cite{Barata:2023qds} at the level of the average transverse momentum distribution obtained from Eq.~\eqref{eq:spec}.

\begin{figure}[htbp]
  \centering
  \includegraphics[width=.45\textwidth]{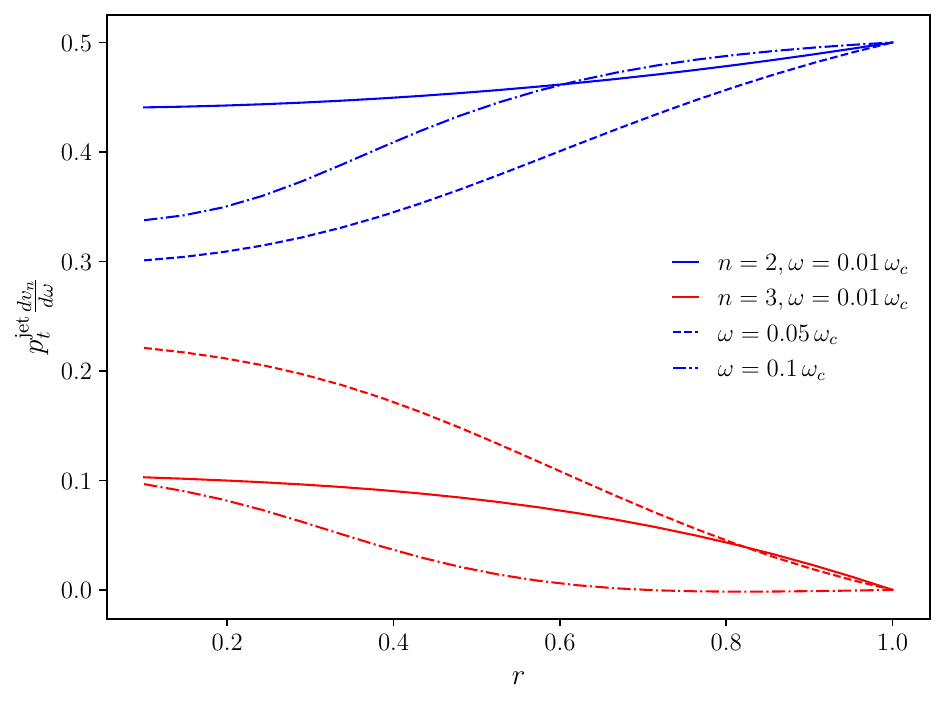}
  \caption{Harmonic decomposition of the energy inside the jet, following Eq.~\eqref{eq:harm} for $\gamma_T=0.5$. We used the same parameters as in Fig.~\ref{fig:JS_1}. }
  \label{fig:harmonics}
\end{figure}

To further illustrate the non-trivial azimuthal dependence of the energy distribution inside the jet, we compute its harmonic decomposition as
\begin{align}\label{eq:harm}
  p_t^{\rm jet}\frac{d v_n(r)}{d\omega} = \int_0^{2\pi} d\alpha \,  p_t^{\rm jet}\frac{d\rho(r)}{d\omega d\alpha} (\cos\alpha)^n\, ,
\end{align}
where $v_n$ denotes the $n$th harmonic.  Notice that unlike in the case of the trigonometric expansion in flow harmonics,  i.e.  in terms of $\cos(n\alpha)$,  here we use a decomposition in terms of $(\cos\alpha)^n$.  These two bases are uniquely related through the multiple-angle trigonometric formulas.  In Fig.~\ref{fig:harmonics} we show the distributions of the harmonics $v_n$ for $n=2,3$. We note that for any odd $n$ the contribution related to the isotropic part of the spectrum $I_0$ vanishes, which explains why the $n=3$ distribution decreases as a function of $r$ and vanishes at $r=1$. Also notice that for larger values of $r$, the shapes in Fig.~\ref{fig:JS_1} become increasingly more isotropic.  Thus, for the odd distributions one should observe a strong ordering in the gluon energies, with the more energetic gluons ordered bottom to top, as is observed in Fig.~\ref{fig:harmonics}.  When going away from this large $r$ region, the ordering of the different gluon energies can change as a consequence of the non-trivial energy  dependence inside the jet.  For instance,  in Fig.~\ref{fig:harmonics} for $n=3$, we observe that the intermediate frequency harmonic is enhanced in the most collinear region. The radial evolution for largest energy is easily understood by looking at the middle column in Fig.~\ref{fig:JS_1}, where one observes that after $r=0.6$ the distribution becomes almost isotropic, thus not contributing to $v_3$. Focusing now on $v_2$, we again note that the $r=1$ point is solely determined by the angular average of $\langle(\cos\alpha)^2/2\pi\rangle_\alpha=1/2$, and thus all gluon energies match. The intermediate region can show non-trivial ordering for the same reason detailed for $v_3$. However, for the even harmonics the distributions should always be strictly growing with the radial distance $r$ due to the fact that the jet shape density becomes increasingly more isotropic. 

%%%%%%%%%%%%%%%%%%
\section{Jet angularities}
\label{sec:jet angularities}
%%%%%%%%%%%%%%%%%%

The integrated jet shape $\rho$ (and its moments) is not an optimal observable to study jet modifications due to medium anisotropies.  Even though the differential jet shape carries non-trivial information about the azimuthal distribution of energy within the jet, such an object can not be experimentally measured in a straightforward manner.  However,  the jet shape is just the first of a family of moments of the energy distribution inside the jet, see for example~\cite{Gallicchio:2010sw,Gallicchio:2012ez, Larkoski:2014pca}.  It is thus natural to study these related distributions, since they can give further information about the modifications to the jet substructure due to matter anisotropies. 

To this end, we consider the radial moments $G_n$ of the energy distribution within a jet, which can be defined by 
\begin{align}\label{eq:moments_jet_shapes}
  G_n = \sum_{i\in {\rm jet}} \frac{p_t^i }{p_t^{\rm jet}} g^{(n)}(r_i)	\, ,
\end{align}
where $g^{(n)}$ is a polynomial corresponding to the particular moment.  The energy carried by $i$th parton in the jet,  located at an angular distance $r_i<R$,  is denoted by $p_t^i$.  We will restrict the discussion to the case of monomial $g^{(n)}$, e.g.  $g^{(0)}\equiv1$ related to the jet shape $\rho$,  $g^{(1)}\equiv r$ corresponding to the jet girth $g$, and $g^{(2)}\equiv r^2$ associated with the squared jet mass $m^2$. 

At LO,  the n$th$ moment,  associated to a particular choice for $g^{(n)}$ in Eq.~\eqref{eq:moments_jet_shapes} is distributed according to
\begin{align}
\label{eq:sigma_LO_new}
  \frac{d\sigma^{\rm LO}}{\sigma d g_n d\alpha } &\equiv 	 \int_0^R d\theta \int_0^1 dx \, \frac{d\sigma}{\sigma d\theta dx d\alpha} \delta(g_n-x\theta^n) \, ,
\end{align}
where $g_n$ corresponds to the girth $g$ for $n=1$, to $m^2$ for $n=2$, and so on.\footnote{Note that $m^2$ here is dimensionless, and should be multiplied by the appropriate power of $p_t^{\rm jet}$ to obtain the physical jet mass.} In the vacuum, the soft and collinear cross-section for the $q\to q+g$ process is given by 
\begin{align}
\frac{d\sigma^{\rm vac}}{\sigma d \theta dx } = \frac{2\alpha_s C_F}{\pi} \frac{1}{x\theta}\, ,
\end{align}
neglecting the phase space where $x\sim1$, which is not captured in Eq.~\eqref{eq:spec} for the medium part. Restricting the discussion to this single channel, Eq.~\eqref{eq:sigma_LO_new} immediately yields
\begin{align}\label{eq:sigma_LO_vac}
	\frac{g_n}{\sigma}\frac{d\sigma^{\rm vac}}{ d g_n d\alpha } = \frac{\alpha_s C_F}{\pi^2\, n } \log \frac{R^n }{g_n}\, .
\end{align}
This fixed order calculation exhibits a logarithmic divergence for small values of $g_n$. As a result, one should resum such terms into a Sudakov factor, for sufficiently small values of $g_n$~\cite{Marzani:2019hun}. Also notice that gradient effects are expected to emerge at smaller values of $g_n$ and, thus, the existence of Sudakov logarithms in the vacuum spectrum can mask the medium effects. It is important to note that the medium contributions will not have such a collinear singularity. Thus, assuming the decomposition valid at LO for the full cross-section
\begin{align}  
  d\sigma^{\rm LO} = d\sigma^{\rm med} +d\sigma^{\rm vac} 	\, ,
\end{align}
we write the LO distribution for $g_n$, including the resummation of the vacuum logarithms to all orders in $\alpha_s$ at leading logarithm accuracy~\cite{Marzani:2019hun}, as
  \begin{align}
    \frac{d\sigma}{\sigma d g_n d\alpha} 	&\approx \frac{d\sigma^{\rm LO }}{\sigma d g_n d\alpha}e^{\int \frac{1}{\sigma}d\sigma^{\rm vac}} \, ,
\end{align}
where the exponential factor corresponds to a cumulative distribution.  The leading term includes the vacuum part given in Eq.~\eqref{eq:sigma_LO_vac}, and also the medium modification in the form of the cross-section
\begin{align}
	\frac{d\sigma^{\rm med}}{\sigma d g_n d\alpha} =  \int_{ \frac{g_n}{R^n}}^1 dx \left[\omega \frac{dI}{d\omega d^2 \k} \frac{(p_t^{\rm jet})^2}{n}  \frac{x^{1-\frac{2}{n}}}{g_n^{1-\frac{2}{n}}}\right]_{\theta^n=\frac{g_n}{x}}\, .
\end{align}
Combining all these elements  we then have the LO form for the medium modified distribution
\begin{align}\label{eq:final_gn_dist}
	\frac{g_n}{\sigma}\frac{d\sigma}{ d g_n d\alpha} &= \Bigg( \int_{ \frac{g_n}{R^n}}^1 dx \, \left[\frac{\omega dI}{d\omega d^2 \k} \frac{(p_t^{\rm jet})^2}{ n }\frac{x\,g_n^{\frac{2}{n}}}{x^{\frac{2}{n}}} \right]_{\theta^n=\frac{g_n}{x}} \nn 
  &\hspace{1cm}+ \frac{\alpha_s C_F}{\pi^2 n} \log \frac{R^n}{g_n} \Bigg) e^{-\frac{\alpha_s C_F}{n\pi} \log^2 \frac{R^n}{g_n}}\, ,
\end{align}
up to an overall normalization factor.  One should notice that under this approximation we neglect the subleading logarithmic terms,  associated with the in-medium part of the normalized cross-section, in the Sudakov factor.  However,  at least a part of these missing terms in the resummation can be recovered,  if we normalize the distribution.

\begin{figure*}[htbp]
  \centering
  \includegraphics[width=.45\textwidth]{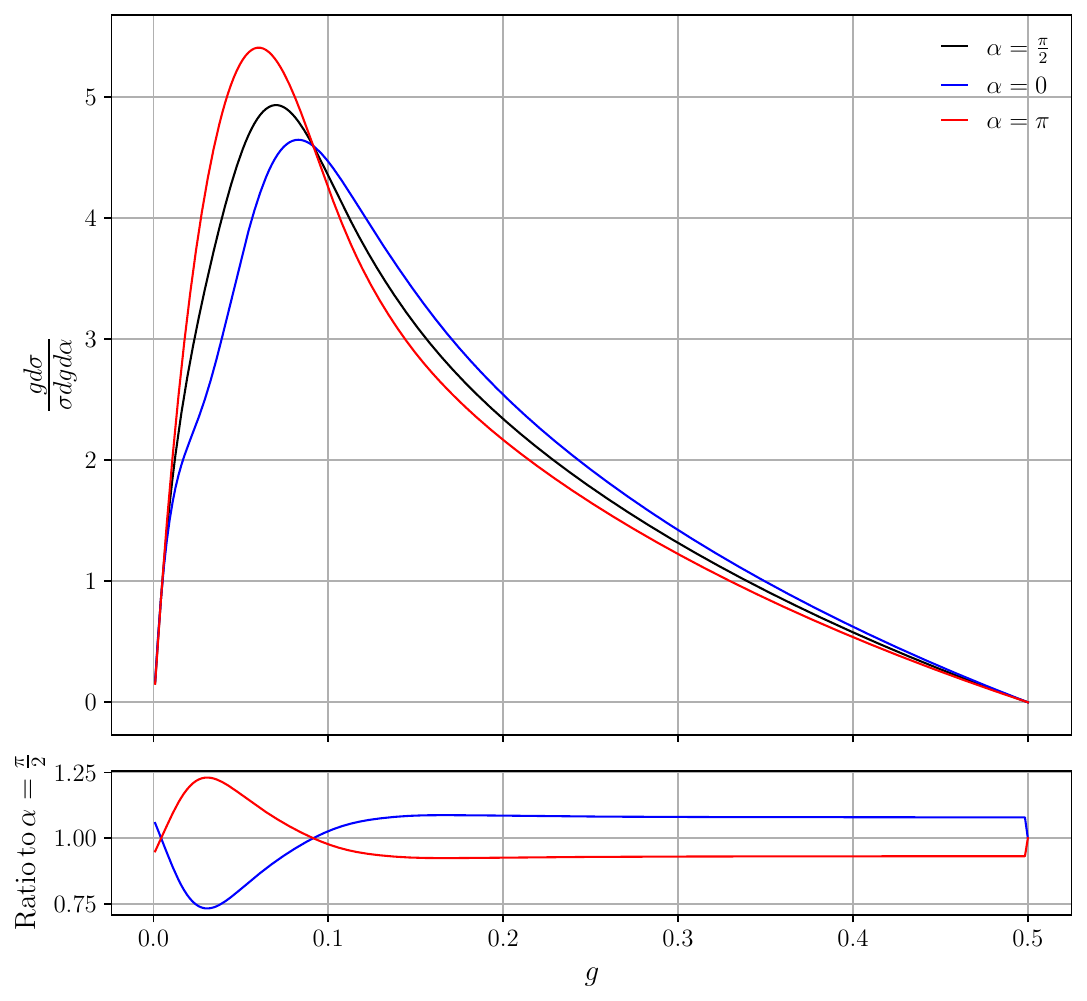}
  \includegraphics[width=.45\textwidth]{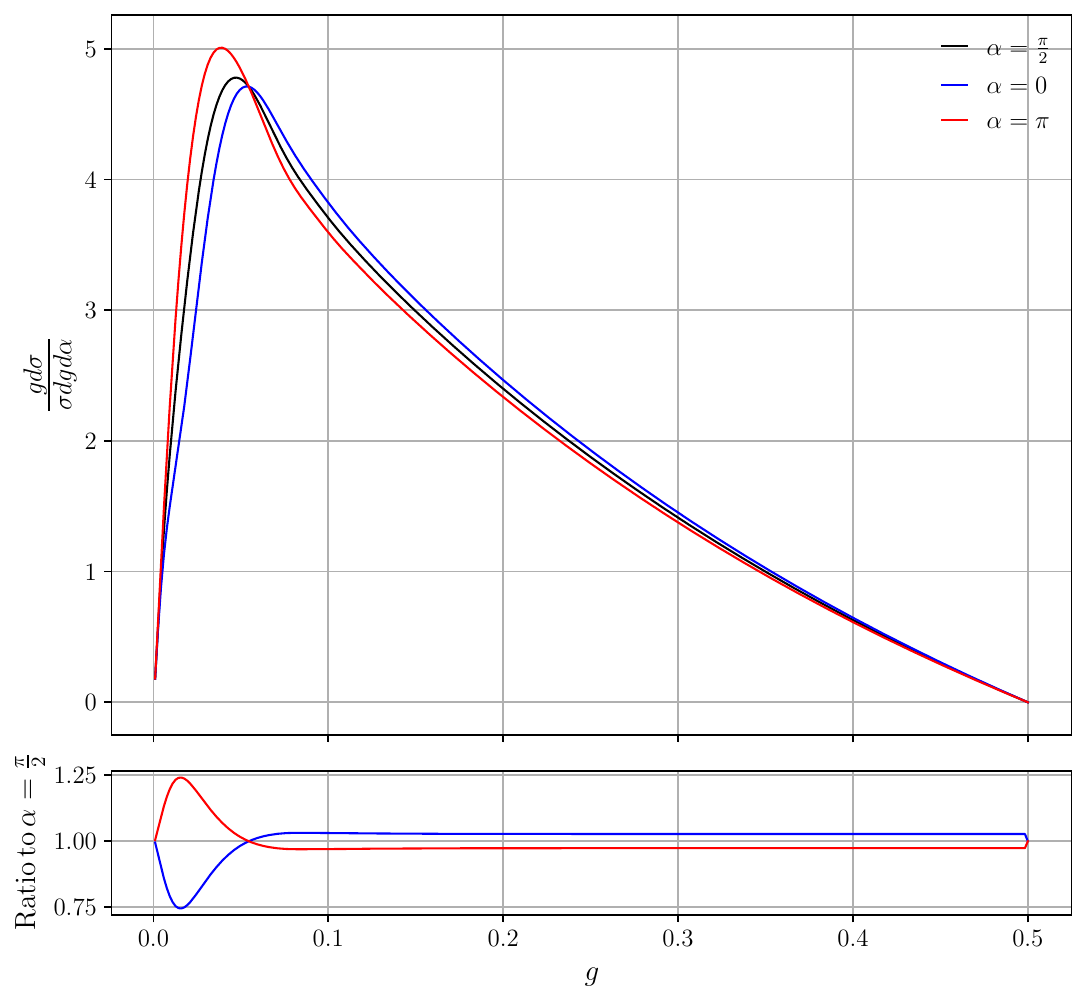}
  \caption{Jet girth distribution for the same parametric settings as used in Fig.~\ref{fig:JS_1},  with $\gamma_T=0.5$, $p_t^{\rm jet}=50$ GeV (left) and $p_t^{\rm jet}=100$ GeV (right). We selected three illustrative values for the azimuthal angles: $\alpha=\pi/2$ (black; no gradient effect), $\alpha=0$ (blue; aligned with gradients) and $\alpha=\pi$ (red; anti-aligned with gradients). The curves on the upper panels are self normalized, while the bottom panels show the ratio to the $\alpha=\pi/2$ curve. Notice that the ratio is taken between the normalized curves. The last point in the ratio plots is set to one.}
  \label{fig:girth}
\end{figure*}

\begin{figure*}[htbp]
  \centering
  \includegraphics[width=.45\textwidth]{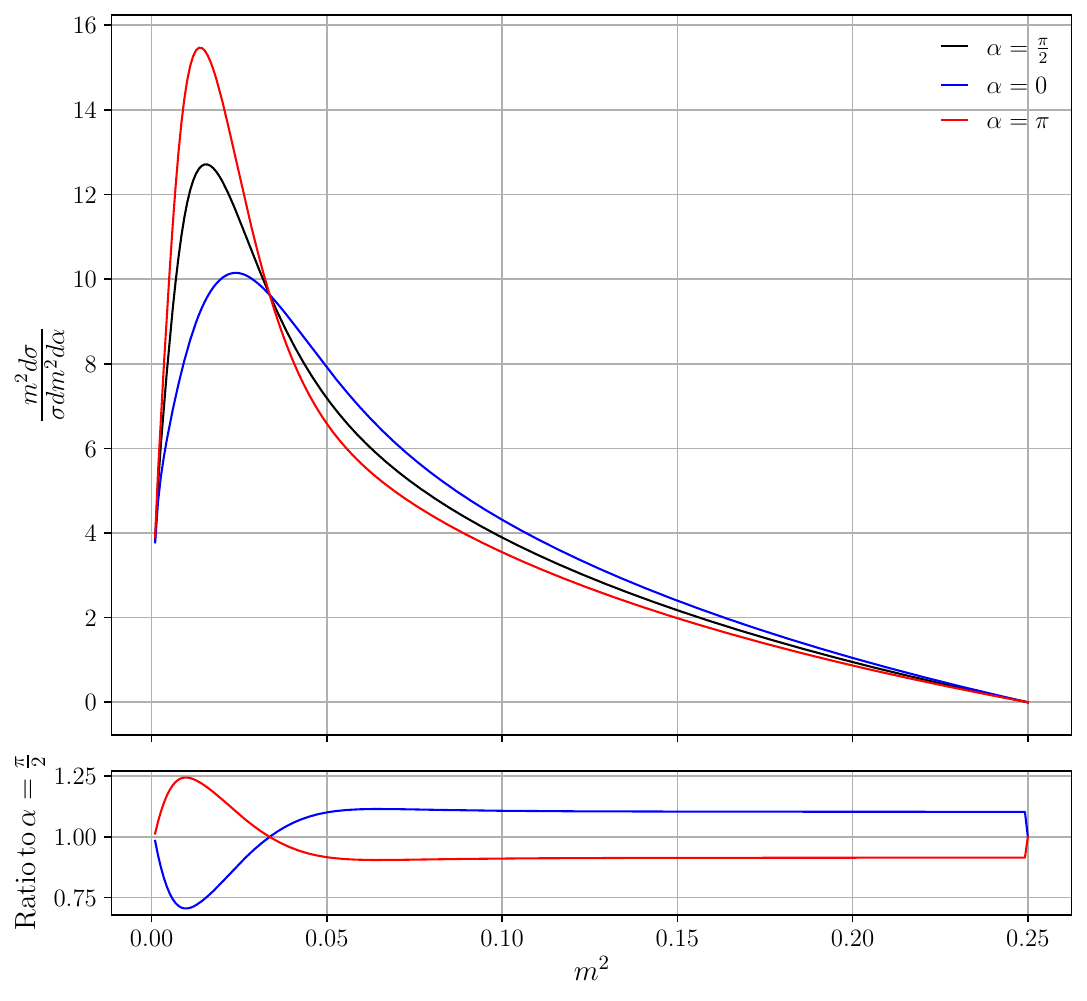}
  \includegraphics[width=.45\textwidth]{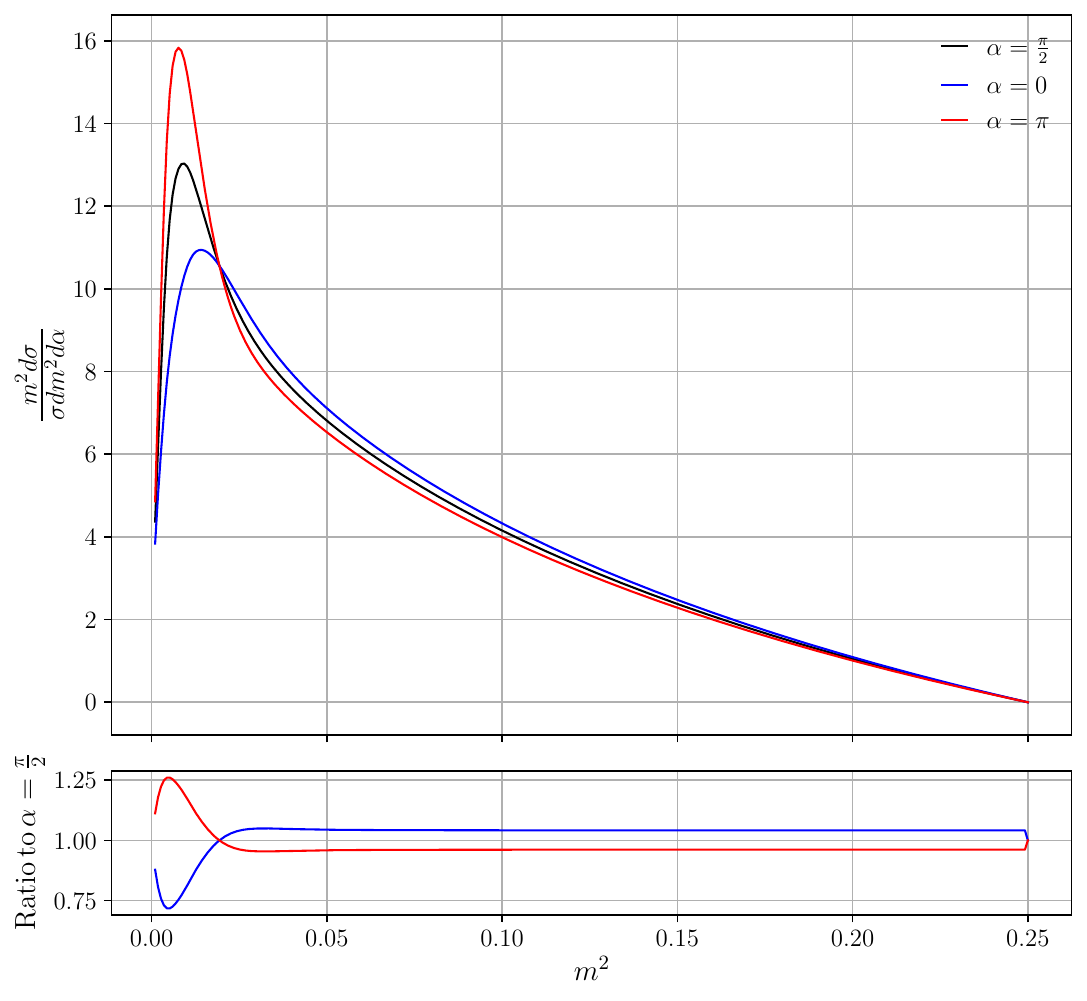}
  \caption{Jet mass distribution for the same parameters and conventions followed in Fig.~\ref{fig:girth}.}
  \label{fig:mass}
\end{figure*}

In Figs.~\ref{fig:girth} and \ref{fig:mass} we show the jet girth and mass distributions computed according to Eq.~\eqref{eq:final_gn_dist}, with $n=1$ and $n=2$ respectively.  All the curves are self normalized.,  i.e.  they are scaled with the average value of the $g_n$.  The right column plots are produced for $p_t^{\rm jet}=100$ GeV, while the left plots take $p_t^{\rm jet}=50$ GeV. As expected, for very energetic jets, the anisotropic corrections become small, since at leading gradient order all corrections are energy suppressed.  We also numerically checked that the contributions along (against) the gradients, i.e. $\alpha=0$ ($\alpha=\pi$), favor larger (smaller) values for the mass/girth and that the distribution's width is wider (narrower) compared to the isotropic QGP scenario.  Finally, we note that larger values of $n$ lead to a better separation between the different curves, but they are also more affected by the Sudakov exponential suppression factor.  These factors play a major role since they suppress much of the softer gluon radiation contributions for the lower value of $g_n$, and thus they constitute a competing (vacuum) effect with respect to the gradient terms. 

The ratios to the isotropic QGP case (bottom plots in Figs.~\ref{fig:girth} and \ref{fig:mass}) are almost symmetric with respect to the unity line,  when comparing $\alpha=0$ and $\alpha=\pi$ lines. This results just from the fact the ratios evolve as $1\pm|\boldsymbol{g}| c(r) + \mathcal{O}(\boldsymbol{g}^2)$, for some function $c(r)$ and where the $\pm$ corresponds to $\alpha=0$ and $\alpha=\pi$, respectively. Also notice that this is only true since we self normalize the plots before taking the ratios.  At large values of $g_n$ the deviations with respect to the isotropic case are roughly sub $10\%$ for all cases, and constitute a  small effect. The corrections around the peak of the distributions can be $\mathcal{O}(25\%)$, and thus could in principle be of phenomenological importance.

%%%%%%%%%%%%%%%%%%
\section{Energy energy correlators }
\label{sec:EEC}
%%%%%%%%%%%%%%%%%%
In the previous sections we have shown that traditional jet shape observables are sensitive to the medium modifications induced in jets by matter gradients. Although the jet shapes studied show a clear azimuthal modulation,
such effects can be either hard to extract experimentally due to contamination from other sources, or masked by competing effects. As a result, it would be desirable to look for other jet observables which could in part mitigate some of these effects and still be sensitive to internal jet scales.

One way to proceed in this direction would require employing jet substructure techniques, see~\cite{Marzani:2019hun} and references there in. These allow, for example, to clean the jet from contaminating sources or pin point interesting structures within it. Although traditional jet shapes can be merged with such techniques, for the azimuthal effects we want to study they only become relevant beyond fixed LO calculations. Instead, we take another route and make use of recent developments using correlations between energy flows inside jets as probes for their internal structure, see e.g.~\cite{Dixon:2019uzg,Chen:2020vvp,Chen:2022swd,Komiske:2022enw,Lee:2022ige}. 
Such correlation functions are particularly interesting due to their ability to resolve internal scales of jets, even though they rely on inclusive inner jet particle distributions. Although the study of these objects is much less developed compared to more traditional jet observables, they posses several theoretical properties which makes their use appealing.

At LO, the only non-trivial object one can compute is the two point correlator, usually referred to as Energy Energy Correlator. In the vacuum, due to spatial isotropy and homogeneity, such an object  can only depend on the absolute value of the spatial separation between the two points where the energy flows are measured. However, in the QGP it is not reasonable to expect that such a large degree of symmetry survives.  Indeed, in our simple matter model the introduction of a spatial gradient results in a preferential direction. As a result, the EEC can depend non-trivially on the polar and azimuthal angles describing the positions of the energy flows on the sphere. Thus, we consider the double differential cumulative distribution\footnote{See also \cite{Kang:2023gvg} for a discussion of similar observables in the context of cold nuclear matter.} 
\begin{align}\label{eq:EEC_def}
  \frac{d \Sigma}{d\theta d\alpha}&= \int d\vec{n}_1 d\vec{n}_2 \,  
 \frac{\langle  \mathcal{E}(\vec{n}_1) \mathcal{E}(\vec{n}_2)\rangle}{(p_{t}^{\rm jet})^2} \delta(\cos(\theta_2-\theta_1)-\cos(\theta)) \nn 
  &\times \delta((\alpha_1-\alpha_2)-\alpha)\nn 
  &= \int dx d\theta_1 d\theta_2 d\alpha_1 d\alpha_2 \, \frac{d\sigma}{\sigma dx d\theta_1 d\alpha_1 d\theta_2 d\alpha_2} x(1-x)   \nn 
  &\times \delta(\cos(\theta_2-\theta_1)-\cos(\theta)) \delta((\alpha_1-\alpha_2)-\alpha)\, ,
\end{align}
where $\langle  \mathcal{E}(\vec{n}_1) \mathcal{E}(\vec{n}_2)\rangle$ denotes the two point correlator of the energy flow operator $\mathcal{E}(\vec{n})$ along the three dimensional unit vector $\vec{n}$, characterized by the angles $(\theta,\alpha)$. The distribution $\Sigma$ corresponds then to a particular projection of the EEC. The second equality in Eq.~\eqref{eq:EEC_def} is valid at LO, after identifying the correlator with the respective energy weighted cross-section. 

In general, simplifying Eq.~\eqref{eq:EEC_def} for arbitrary matter geometries beyond this point is not possible using analytical methods. However, in our simple scenario, where the anisotropies are encapsulated by the vector $\boldsymbol{g}$, one can easily show that the EEC reduces to
\begin{align}\label{eq:final_EEC}
  \frac{d \Sigma}{d\theta d\alpha} 
  &= \int_0^1 dx \Biggl( \frac{\alpha_s C_F}{\pi^2} \frac{1}{x\theta}+ \omega \frac{ dI}{d\omega d^2 \k} p_t^{\rm jet} k  \Biggr)x(1-x)  \, ,
\end{align}
where we have used the results of the previous section to include the vacuum and medium pieces and one should take the limit $x\to0$. At this point we should notice a conceptual inconsistency in our calculation: we are integrating the medium-induced spectrum in the domain $0<x<1$ in Eq.~\eqref{eq:final_EEC}, while the spectrum is formally derived in the limit $x\ll 1$. As mentioned in the introduction, it is theoretically challenging to lift this approximation. However, we note that even using this spectrum, the resulting EEC in isotropic matter has features qualitatively similar to calculation performed for finite energy gluons, for more details see~\cite{Andres:2022ovj, Andres:2023xwr,Barata_HP}. 

\begin{figure*}[htbp]
  \centering
  \includegraphics[width=.45\textwidth]{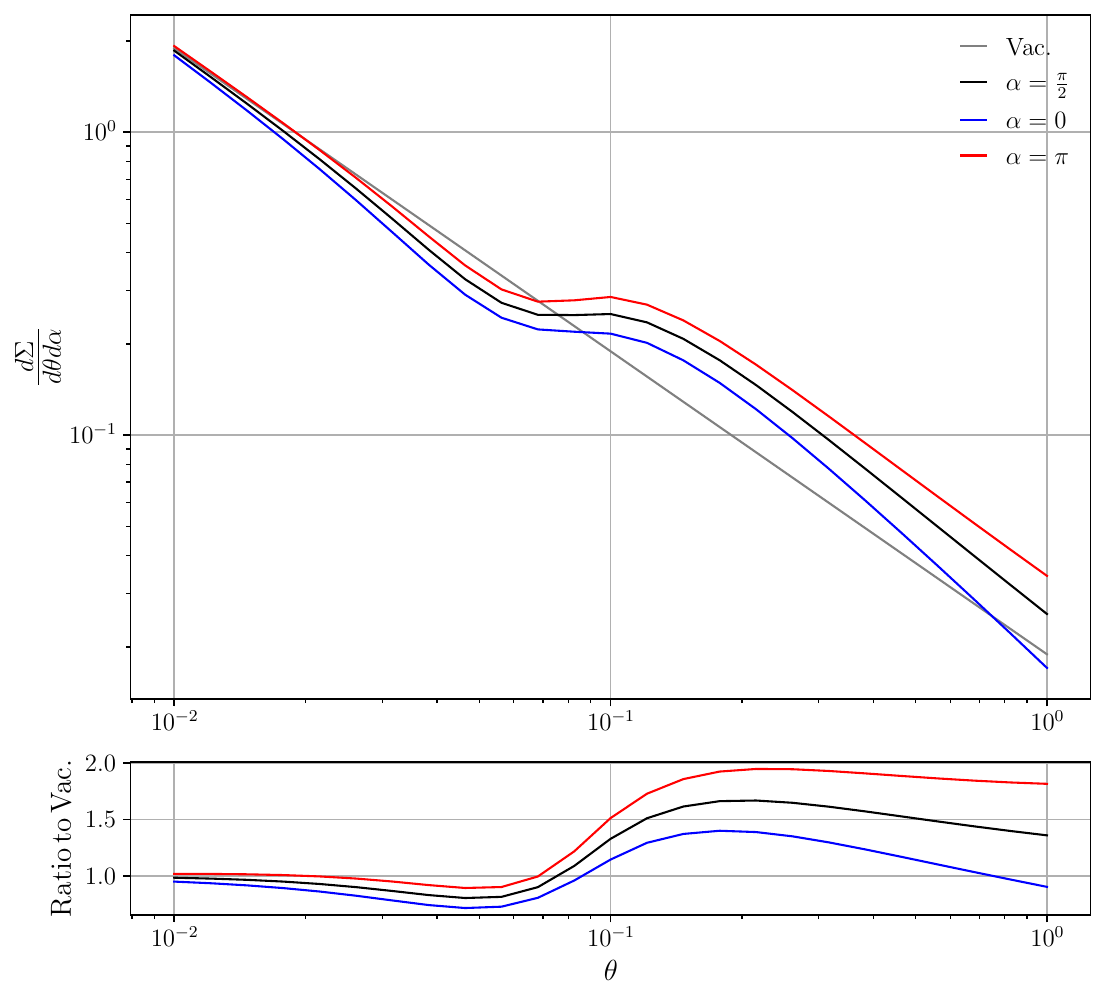}
  \includegraphics[width=.45\textwidth]{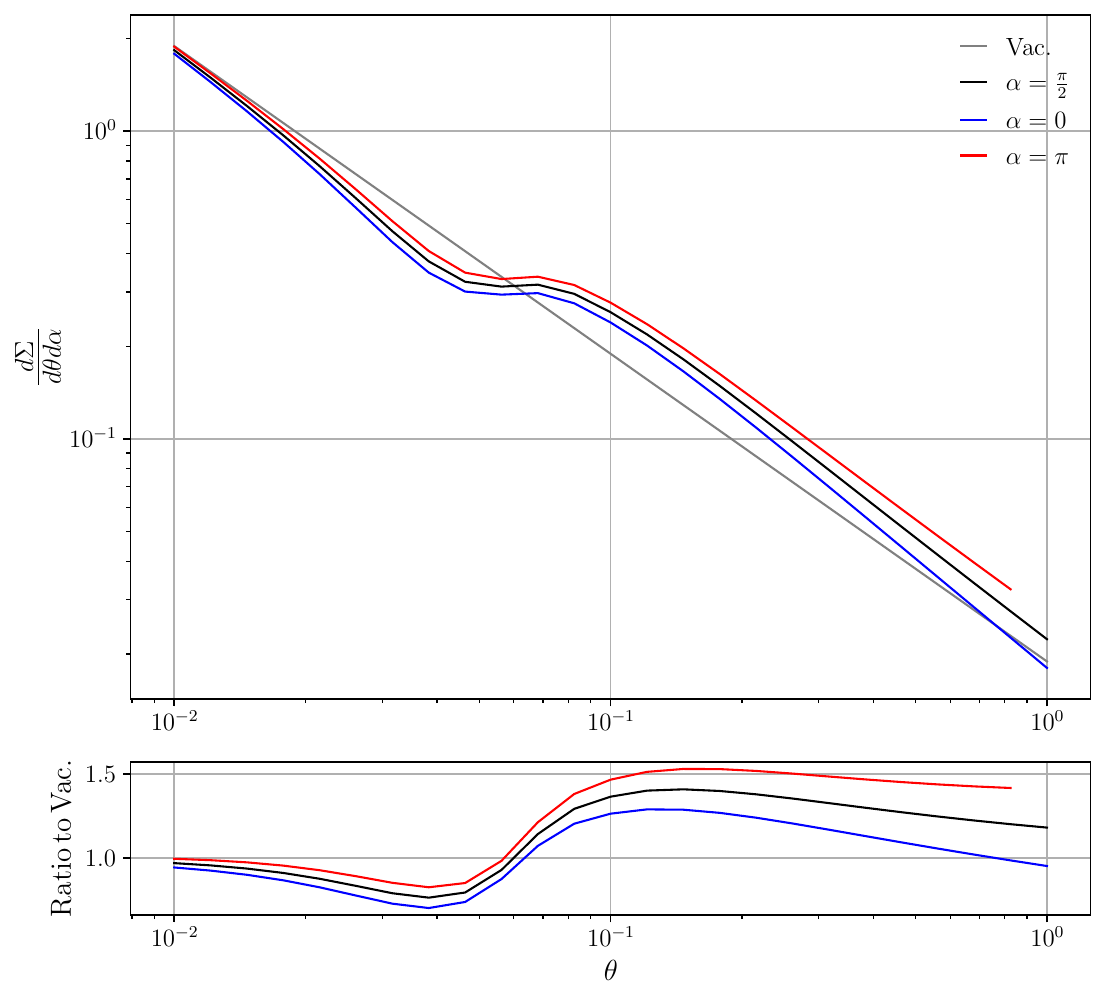}
  \caption{EEC double differential distribution, following the parameters and conventions used in Fig.~\ref{fig:girth}, with the left plot using $p_t^{\rm jet}=50$ GeV and $p_t^{\rm jet}=100$ GeV on the right. The pure vacuum distribution is additionally plotted in gray, for better visualization of the purely medium-induced contributions. On the right hand side plot, the $\alpha=\pi$ does not cover the full angular domain purely due to numerical artifacts.}
  \label{fig:eec}
\end{figure*}

The numerical evaluation of Eq.~\eqref{eq:final_EEC} is shown in Fig.~\ref{fig:eec}, for the same values of $p_t^{\rm jet}$ considered in Fig.~\ref{fig:girth} and following the same conventions. The vacuum curves evolve as $1/\theta$ at LO. The isotropic matter curves (black) display an enhancement around a particular angle; this is related to the existence of a characteristic angular scale controlling the emission of gluons in a dense QCD medium, see e.g.~\cite{Mehtar-Tani:2013pia} for further discussion. Such a feature is qualitatively similar to the one observed for calculations in dense matter beyond the small $x$ limit considered here~\cite{Andres:2022ovj,Andres:2023xwr}. As in the other cases, for larger jet energies, the matter anisotropy effects are smaller. However, one appealing feature suggested by the EEC calculations is the persistent separation between the different curves over a reasonably large domain in $\theta$.  Of course, at very small values of the angular separation, above the region sensitive to non-perturbative physics, the collinear effects are expected to dominate over all the medium effects, and thus there is an effective lower bound for this observation. We also want to note that for higher order calculations, the shape of this distribution could likely be heavily modified. However, we still would expect that there to be a remaining azimuthal dependence, regardless of the behavior in $\theta$. To test this claims requires performing more accurate calculations, which we leave for the future work.

\section{Conclusion}\label{sec:conclusion}
We have presented an exploratory study of hydrodynamic matter gradient effects on several jet observables. Our calculations are done at leading order in $\alpha_s$ and assuming the production of soft induced gluon radiation from a quark hard source.  The medium is modeled as a static brick of matter, with an anisotropic direction,  defined by the hydrodynamic matter gradients. We decide to put the emphasis on computing jet observables in the regime where the theoretical results used are well understood and under control.

We have found that the presence of the anisotropy leads to a non-trivial distribution of matter inside the jet, which can be characterized by the computation of the jet shape density.  This energy distribution can be further characterized in terms of its harmonic decomposition,  which now exhibits odd terms,  absent for isotropic matter.  However,  these observables are not optimal to extract information from the jet,  since they have smaller sensitivity to the inner structure and are easily contaminated by radiation coming from uncorrelated sources.  As a result,  we then considered the lowest order jet angularities,  i.e.  jet girth and mass.  We showed that on these observables,  which are sensitive to the jet substructure,  matter anisotropies lead to a shift towards larger (smaller) values of the distribution when measured along (against) the dominant anisotropy direction.  On top of that,  the width of the distribution is also modified differently according to the azimuthal direction.  However,  vacuum Sudakov effects compete with the anisotropic effects in the regions where these are more dominant.  As a result,  the observation of these effects is not straightforward.  One way to surpass this shortcoming is to consider jet observables which can look more differentially inside of the jet,  using modern jet substructure techniques.  However,  most of these resources only become interesting when going beyond the LO calculation we perform.  We instead consider the behavior of EECs measured in jets (although,  using the oversimplified soft limit for the medium modified cross-section),  which have been argued to provide information about the inner jet scales.  Indeed,  we observe that these are also sensitive to the matter anisotropies over a large angular region. For any further analysis, this consideration of the EECs should be completed with the use of a consistent form for the medium-induced spectrum applicable in the whole relevant kinematic region,  and we leave that for future work.

\section*{Acknowledgements}
We are grateful to Carlota Andrés, Fabio Dominguez, Jack Holguin, and Ian Moult for clarifications on~\cite{Andres:2022ovj}. JB is also grateful to Yacine Mehtar-Tani, Alba Soto-Ontoso, and Robert Szafron for parallel discussions.  JB was supported by the U.S. Department of Energy under contract DE-SC0012704. AVS and JGM were partly supported by the European Research Council project ERC-2018-ADG-835105 YoctoLHC.   The work of AVS is also supported by the Marie Sklodowska-Curie Individual Fellowship under JetT project (project reference 101032858).  AVS also acknowledges further support from the Maria de Maetzu excellence program under projects CEX2020-001035-M; from the Spanish Research State Agency under project PID2020-119632GB-I00; from Xunta de Galicia (Centro singular de investigación de Galicia accreditation 2019-2022), and from European Union ERDF.  JGM was also supported by Fundação para a Ciência e a Tecnologia (FCT), I.P., project CERN/FIS-PAR/0032/2021.

\appendix

\bibliographystyle{apsrev4-1}
\bibliography{references.bib}

\end{document}